\documentclass[reprint,amsmath,amssymb,aps,pra,longbibliography,nofootinbib]{revtex4-1}
\usepackage{float}
\usepackage{graphicx}
\usepackage{dcolumn}
\usepackage{bm}
\usepackage{multirow}
\newcommand\Tstrut{\rule{0pt}{1ex}}         
\newcommand\Bstrut{\rule[1ex]{0pt}{0pt}}
\usepackage[%
  colorlinks=true,
  urlcolor=blue,
  linkcolor=blue,
  citecolor=blue
]{hyperref}

\begin{document}
\title{Isotope Shifts in the 7s$\rightarrow$8s Transition of Francium: Measurements and Comparison to \textit{ab initio} Theory}
\author{M. R. Kalita, J. A. Behr, A. Gorelov  and M. R. Pearson }
\affiliation{%
 TRIUMF, Vancouver, Canada BC V6T 2A3
}%
\author{A. C. DeHart, G. Gwinner and M. J. Kossin}
\affiliation{
 Department of Physics and Astronomy, University of Manitoba, Winnipeg, Canada MB R3T 2N2
}%
\author{L. A. Orozco}
\affiliation{
 JQI, Department of Physics and NIST, University of Maryland,College Park, Maryland 20742, USA
}%
\author{S. Aubin}
\affiliation{
  Department of Physics, College of William and Mary, Williamsburg, Virginia 23186, USA
}%
\author{E. Gomez}
\affiliation{
  Instituto de F\' {i}sica, Universidad Aut\'{o}noma de San Luis Potos\'{i}, San Luis Potosi 78290, Mexico
}%
\author{M. S. Safronova}
\affiliation{
  Department of Physics and Astronomy, U.
Delaware, Newark, Delaware 19716, USA
}%
\affiliation{
 JQI, Department of Physics and NIST, University of Maryland, College Park, Maryland 20742, USA
}%
\author{V. A. Dzuba and V. V. Flambaum}
\affiliation{School of Physics, U. New South Wales, Sydney
2052, Australia
}%



\begin{abstract}
We observe the electric-dipole forbidden $7s\rightarrow8s$ transition in the francium isotopes  $^{208-211}$Fr and  $^{213}$Fr using a two-photon excitation scheme. We collect the atoms online from an accelerator and confine them in a magneto optical trap for the measurements. In combination with previous measurements of the $7s\rightarrow7p_{1/2}$ transition we perform a King Plot analysis. We compare the thus determined ratio of the field shift constants (1.230 $\pm$ 0.019)  to results obtained from new \textit{ab initio} calculations (1.234 $\pm$ 0.010) and find excellent agreement.

\end{abstract}

\pacs{Valid PACS appear here}
\maketitle


\section{\label{sec:level1}Introduction}
The isotope shift in the transition energies of an atom arises due to a combination of nuclear and atomic effects. It is an important benchmark, as it can provide information about the nuclear charge distribution and its change as more neutrons are added; the shift also depends on electron correlations.
The FrPNC collaboration at TRIUMF has been  studying francium with the ultimate goal of measuring atomic parity non-conservation (APNC)     \cite{Gwinner2007,PhysRevA.75.033418}. Others have also proposed to use francium for APNC studies \cite{ATUTOV2004421}, and to search for time reversal violation through the existence of a permanent electric dipole moment (EDM) of the electron \cite{PhysRevX.2.041009, Inoue2015}.
These proposals require quantitative understanding of the atomic and nuclear structure, and in particular the overlap of the electronic wave functions with the nucleus. This overlap can be tested by comparing the measurements of  hyperfine structure and isotope shift  in chains of isotopes  to the {\it ab initio} calculations   \cite{isotopebook}.

Testing the accuracy of the {\it ab initio} theory for  field shifts in heavy
atoms is also crucial for  extraction of the change of nuclear radii in
 superheavy elements \cite{LaaLauBac16}. Combining theoretical and experimental
isotope shift values allows the extraction of the differences in the nuclear radii of these atoms and provides an insight into their nuclear structure.
Studies of isotope shift of superheavy elements  are also of  interest for astrophysics \cite{DzuFlaWeb17}.
All of these projects require reliable benchmarks of theoretical calculations in order to verify  their theory uncertainties.
 Measurements of the field shift ratios for different atomic transitions  are of particular interest owing to the recently found disagreement of the Ca$^+$ D1/D2 field shift measurement with all theoretical predictions \cite{ShiGebGor17}.  Isotope shift measurements have also been proposed  as a new method to probe new light force-mediators \cite{BerBudDel17}.

Here, we report the observation of the electric dipole forbidden $7s\rightarrow8s$ atomic transition in the francium isotopes  $^{208-211}$Fr and  $^{213}$Fr using a single-frequency two-photon excitation scheme, its isotope shift and the comparison to {\it ab initio} theory. This transition in francium is of particular interest for APNC experiments, as it is electric-dipole forbidden by electromagnetism but slightly allowed by the weak interaction. The landmark APNC experiments in cesium performed in Paris and Boulder used the equivalent  $6s\rightarrow7s$ transition in cesium {\cite{PhysRevLett.82.2484,doi:10.1139/p99-002, BOUCHIAT1982358}}.
 Our isotope shift measurements are complementary to  hyperfine splitting measurements, which also depend on the electronic wave functions at the nucleus. Together with  the information obtained from the change in the nuclear magnetization from the measurements of hyperfine anomalies allows to create a better picture of the nuclear structure \cite{PhysRevLett.83.935, PhysRevLett.115.042501}. In contrast,  measurements of the electronic dipole matrix elements (obtained from lifetime measurements of excited atomic states) probe the wave functions predominantly at large distances from the nucleus \cite{eduardoRev}. 

We divide this paper into the following sections: In section \ref{Theory} we briefly discuss the general theory relevant to the measured isotope shifts, in section \ref{Field shift theory} we present the theoretical calculations of the field shift, in section \ref{Experimental details} we describe the experimental details,  in section \ref{Results} we discuss our experimental results, section \ref{King Plot analysis} contains a King Plot analysis and the comparison with the theoretical predictions, closing with conclusions in  section \ref{Conclusion}.
\section{\label{sec:level1}Theory}\label{Theory}

Single-photon electronic transitions between states of same parity in atoms are forbidden by electric-dipole selection rules; however, a two-photon transition is allowed between states of the same parity. The selection rules for a two-photon transition where both photons are far off resonance from any intermediate states are  $\Delta F= 0$ and $\Delta m_F= 0$ \cite{581793936}.

Using two-photon spectroscopy in our set-up and previously measured hyperfine splittings of the $7s$ and $8s$ states we  obtain the center of gravity (C.O.G) of the $7s\rightarrow8s$  transition  in five different isotopes of francium that we collect online from an accelerator  and capture in a magneto-optical trap (MOT). From these measurements we   deduce the isotope shifts and perform a King Plot analysis \cite{King:63}.  Optical isotope shifts are discussed in detail in Refs. \cite{HEILIG1974613,isotopebook}. Here we briefly review the  theory that is relevant to the measurements reported in this paper.

For heavy elements the optical isotope shift $\delta\nu^{A A^{'}}_{IS}$, between isotopes  with mass number $A$ and $A^{'}$ and nuclear mass $M_A$ and $M_{A^{'}}$ respectively, can be written as\footnote{In the relativistic case,  $F \delta \langle r^2\rangle$ is  replaced by
$\tilde{F} \delta \langle r^{2\gamma}\rangle$, where $\gamma=(1- Z^2 \alpha^2)^{1/2}$ \cite{DzuFlaWeb17}.}
\begin{equation}
\delta\nu^{A A^{'}}_{IS}=(N+\mathcal{S})\frac{M_A-M_{A^{'}}}{M_AM_{A^{'}}}+F\delta\langle r^2 \rangle^{AA^{'}}.
\label{eq:shift}
\end{equation}

$N$ is the normal mass shift (NMS) constant, and $\mathcal{S}$ is the specific mass shift (SMS)
constant that stems from the changing mass of the nucleus between
isotopes. The contribution of the normal mass shift to the frequency of
an optical transition can be written in the non-relativistic limit as
\begin{equation}
\delta \nu_{NMS}^{AA^{'}}=\nu({A}^{'})\frac{m_e(M_A-M_{A^{'}})}{M_{{A}^{'}}(M_A+m_e)},
\end{equation}
  where $m_e$ is the mass of the electron and $\nu(A^{'})$ is the transition frequency of an isotope with mass number $A^{'}$.

The specific mass shift is  hard to calculate accurately owing to
poor convergence of the perturbation theory for this quantity.
This issue has been discussed in detail in \cite{SafJoh08}. However, the contribution
of the mass shift (both normal and specific) is small for heavy atoms and
simple estimations should be sufficient. Moreover, an earlier study of francium
isotope shifts has demonstrated that NMS and SMS strongly cancel each other
and the residual is at the level of the accuracy of the theoretical field
shift calculations \cite{SafJoh08}.

In the traditional approach, $F$ is the field shift constant that takes into
account the modification of the Coulomb potential of a point-charge by that
of the finite size of a nucleus. 
However, $F$ also depends on the nuclear radius, and
this dependence may be large for heavy atoms.
Nevertheless, if we consider neighbouring isotopes with small
differences between mass numbers, the dependence of $F$ on
the nuclear radius between these isotopes can be neglected.
We  check this for the francium isotopes considered in this work.

$F$ is  a relatively simple single-electron scalar operator.
Unlike $\mathcal{S}$, which is a two-electron operator of rank one, the
field shift can be more easily included into the available, accurate, \textit{ab initio} atomic
methods. In this work, we use two completely different theory methods that we 
describe in Sec.~\ref{sec:th} and compare the results for the field shift values  to evaluate the theoretical uncertainty.

The values of the quantities $N$, $\mathcal{S}$ and $F$  as defined  are specific to a particular electronic transition in an atom.
In our experiment, we obtain the total isotope shift $\delta\nu^{A A^{'}}_{IS}$ for the $7s\rightarrow8s$ transition as expressed by Eq. \ref{eq:shift}.

\section{\label{sec:th}Theoretical calculations  of the field shifts }\label{Field shift theory}
\subsection{The all-order method}
We use a linearized variant of the relativistic coupled-cluster method with single, double, and partial triple
excitations \cite{SafJoh08}, which is referred to as the all-order method.
The exact many-body wave function in the coupled-cluster method is represented in the form
\cite{CK:60}
\begin{equation}
|\Psi \rangle = \exp(S) |\Psi^{(0)}\rangle, \label{cc}
\end{equation}
where $|\Psi^{(0)}\rangle$ is the lowest-order atomic state vector.
The operator ${S}$ for an N-electron atom consists of
``cluster'' contributions from one-electron, two-electron, $\cdots$,
N-electron excitations of the lowest-order state vector $|\Psi^{(0)}\rangle$. Expanding the exponential in Eq.~(\ref{cc}) in terms of the $n$-body excitations ${S}_n$,
and limiting the expansion to terms linear in single, double, and
valence triple contribution, we get the wave function
of a monovalent atom in state $v$:
\begin{eqnarray}
&&|\Psi_v \rangle = \left\{ 1+S_1+S_2+S_{3v} \right\}|\Psi_v^{(0)}\rangle\\
&&                = \left[ 1 + \sum_{ma} \, \rho_{ma} a^\dagger_m a_a + \frac{1}{2} \sum_{mnab} \rho_{mnab} a^\dagger_m a^\dagger_n a_b a_a + \right.   \nonumber \\
&& +  \sum_{m \neq v} \rho_{mv} a^\dagger_m a_v + \sum_{mna} \rho_{mnva} a^\dagger_m a^\dagger_n a_a a_v \nonumber \\
&&+ \left. \frac{1}{6} \sum_{mnrab} \rho_{mnrvab}
 a_m^{\dagger } a_n^{\dagger} a_r^{\dagger } a_b a_aa_v \right]|
\Psi^{(0)}_v \rangle, \label{eq1}
\end{eqnarray}
where $|\Psi_v^{(0)}\rangle$ is the lowest-order atomic state vector.
In Eq.~(\ref{eq1}), $a^\dagger_i$ and $a_i$ are creation and annihilation operators for
an electron state $i$, the indices $m$ and $n$ range over all possible virtual states while
 indices $a$ and $b$ range over all occupied core states. The quantities $\rho$
are excitation coefficients.
The  single double (SD) method is the linearized coupled-cluster method restricted to
single and double excitations only.
The all-order singles-doubles-partial triples (SDpT) method is an extension of
the SD method in which the dominant part of $S_{3v}$  is treated perturbatively.
A detailed description of the SDpT method is given in Ref.~\cite{SafJoh08}.

 To derive equations for the excitation coefficients,
 the wave function $|\Psi_v\rangle$ is substituted into
 the  many-body Schr\"{o}dinger equation
$H | \Psi_v\rangle=E| \Psi_v\rangle, \label{eq2}$
and terms on the left- and right-hand sides are matched,
based on the number and type of operators they contain,
giving the equations for the excitation coefficients.
The Dirac-Hartree-Fock (DHF) starting potential with the inclusion of the Breit interaction is used to produce a finite basis set of the
orbitals for all subsequent calculations.
The equations for the excitation coefficients are solved  iteratively until the valence correlation energy converges to
 a specified numerical accuracy. This procedure effectively sums the series of the dominant many-body perturbation theory (MBPT) terms, with each iteration picking  up a new order of MBPT. Thus, the method includes dominant correlation corrections to all orders of  MBPT.

\subsection{Field-shift calculations: Method I}

If we describe the nucleus as a uniformly charged ball of radius $R$,
the change in the nuclear potential induced by a change in the nuclear radius $\delta R$,
is given by
\begin{equation}
\delta V(r) = \frac{3Z}{2R^2}\left[ 1- \frac{r^2}{R^2} \right] \delta R,
\end{equation}
$r\leq R$. Re-writing this result in terms of the mean square radius $\langle r^2 \rangle=(3/5) R^2$, we define a field-shift
operator $F(r)$ as \cite{WJbook}
\begin{equation}
\delta V= F(r) \delta \langle r^2 \rangle,
\end{equation}
\begin{eqnarray}
F(r)& = & \frac{5Z}{4R^3}\left[ 1- \frac{r^2}{R^2} \right], r\leq R \nonumber \\
    & = & 0, r > R.
\end{eqnarray}

When we use a more elaborate Fermi distribution to describe the nucleus
\begin{equation}
\rho(r)=\frac{\rho_0}{1+\mathrm{exp}([r-c]/a)},
\end{equation}
where $c$ is the 50\% fall-off radius of the density, and $a$ is related to the 90\%-10\% fall-off distance by $t=4\mathrm{ln}(3)a$,
we find negligible differences with the results obtained using the formula for a simple uniform ball for the field-shift operator and
a variant that uses the Fermi distribution.

In our first method we use the all-order approach and we calculate the field shift constant as an expectation value of the field-shift operator $\langle F \rangle$ given by
\begin{equation}
F_{wv}=\frac{\langle \Psi_w |F| \Psi_v \rangle}{\sqrt{\langle \Psi_v
| \Psi_v \rangle \langle \Psi_w | \Psi_w \rangle}}, \label{eqr}
\end{equation}
 where $|\Psi_v\rangle$  and $|\Psi_w\rangle$ are given by the
expansion (\ref{eq1}) limited to single and double excitations.
 The resulting expression for the numerator of
Eq.~(\ref{eqr}) consists of the sum of the DHF matrix element
$f_{wv}$ and twenty other terms that are linear or quadratic
functions of the excitation coefficients.

\begin{table*}[t]
\caption{\label{tab:theory} Field shift constants $F$ (in MHz/fm$^2$) of francium levels calculated using different methods.  ${\mathcal{R}}$ is the ratio of the field shift constants for the $7s\rightarrow7p_{1/2}$ and $7s\rightarrow8s$ transitions defined by Eq.~(\ref{eqR}).
Approximations: DHF$\rightarrow$lowest order Dirac-Hartree-Fock, MBPT 2$\rightarrow$ second-order many-body perturbation theory,
MBPT 3 $\rightarrow$ third-order many-body perturbation theory, All-order - linearized coupled-cluster method with single-double (SD) and partial triple (SDpT) excitation; SD$_{\textrm{sc}}$ and SDpT$_{\textrm{sc}}$ are scaled results that include estimates of the dominant higher excitations. BO+fit  results are obtained using the Brueckner orbitals with fitting to experimental energies. }
\begin{ruledtabular}
\begin{tabular}{llccccc}
      \multicolumn{1}{c}{Method} &
      \multicolumn{1}{c}{Approximation} &
      \multicolumn{1}{c}{$F$($7s$)} &
        \multicolumn{1}{c}{$F$($8s$)} &
            \multicolumn{1}{c}{$F$($7p_{1/2}$)} &
  \multicolumn{1}{c}{$F$($7p_{3/2}$)} &
      \multicolumn{1}{c}{${\mathcal{R}}$}   \\
      \hline
 I	&	DHF	            &	-14239	&	-3649	&	-485	&	0	&	1.299\\
 I	&	All-order SD	&	-22522	&	-4677	&	-683	&	348	&	1.224\\
 I	&	All-order SDpT	&	-21268	&	-4554	&	-674	&	304	&	1.232\\
 I	&	All-order SD$_{\textrm{sc}}$	&	-21647	&	-4602	&	-670	&	333	&	1.231\\
 I	&	All-order SDpT$_{\textrm{sc}}$&	-21618	&	-4603	&	-687	&	312	&	1.230\\
 II	&	MBPT 2	        &	-22480	&	-4732	&	-695	&	311	&	1.227\\
 II	&	MBPT 3	        &	-19441	&	-4359	&	-455	&	449	&	1.259\\
 II	&	BO+fit	        &	-20947	&	-4381	&	-670	&	310	&	1.224\\
 II	&	All-order SD	&	-21236	&	-4436	&	-675	&	333	&	1.224\\
 II	&	All-order SD+E3&	-20181	&	-4322	&	-599	&	371	&	1.235\\
 II	&	All-order SDpT	&	-20582	&	-4421	&	-635	&	338	&	1.234\\
 II	&	Final 	        &-20580(650)&-4420(100)	&	-635(40)&338(33)&	1.234(10)\\
\end{tabular}
\end{ruledtabular}
\label{tab:1}
\end{table*}

\subsection{Field-shift calculations: method II}

In the second method,  we use an
all-order finite-field approach \cite{DzuJohSaf05}.
Calculations of the field shift are done for the reference isotope A with a nuclear charge radius $R$ by replacing
a nuclear potential $V(r)$  by
\begin{equation}
V(r) +\lambda \delta V(r)
\end{equation}
where $\lambda$ is a scaling parameter
\begin{equation}
\delta V(r) = \frac{dV}{dR} \delta \langle R \rangle.
\end{equation}
The Fermi distribution is used for the charge distribution and the derivative
$\frac{dV}{dR}$ is calculated numerically.
 The value of $\lambda$ is chosen in
such a way that the corresponding change in the nuclear potential is
sufficiently small for the final energy
to be a linear function of $\lambda$ but
much larger than the numerical uncertainty of the
calculations. The calculations are carried out for several values of
$\lambda$ and the field shift constant for an
atomic state $v$ is calculated as a derivative
\begin{equation}
F=\frac{dE_v(\lambda)}{d\lambda}.
\end{equation}
Therefore, the calculation of the field shift constants reduces to the calculation of the energy in this method.
\subsection{Theory results and discussion}

The results for the field shift constants $F$ of francium levels calculated using both methods are given in Table~\ref{tab:theory}.
 ${\mathcal{R}}$ is the ratio of the field shift constants for the $7s\rightarrow7p_{1/2}$ and $7s\rightarrow8s$ transitions:
 \begin{equation}
 {\mathcal{R}}=\frac{F(7p_{1/2})-F(7s)}{F(8s)-F(7s)}.
 \label{eqR}
 \end{equation}
 Results obtained in several approximations are given  for both methods. The
DHF lowest order  matrix elements are given to show the size of the correlation corrections. The all-order  single-double (SD) and partial triple (SDpT) results are listed in the SD and SDpT rows. In method I, some classes of omitted contributions from
higher excitations may be estimated by the scaling procedure described in \cite{SafJoh08}, these results are listed with the subscript ``sc''. For method II, we also include the field shift constants obtained using the second- and third-order MBPT energy calculations to show the size of the third and higher-order corrections. The energy in the SD approximation is missing a part of the third-order contribution, which is restored in the results in the ``SD+E3'' row. The SDpT energies include a complete third-order contribution and do not need to be corrected. We also carried out other calculation using Brueckner orbitals (BO) with fitting of the correlation potential to the experimental energies, described in \cite{DzuJohSaf05}. The results are listed in the row labelled ``BO+fit''.
We take the \textit{ab initio} method II SDpT results as final. The uncertainties are estimated from the spread of the all-order results.
We note that the uncertainty of the $7s$ field shift constant was underestimated in \cite{DzuJohSaf05}.
The relative uncertainty in the ratio ${\mathcal{R}}$ is smaller than the uncertainties in the field shift constants for each level as
correlation corrections to the $7s$ and $8s$ states are similar, and the field shift for the $7p_{1/2}$ level is small in comparison to the field shifts of the $7s$ and $8s$ levels.


\section{\label{sec:level1}Experimental details}\label{Experimental details}
We use francium ions produced at the Isotope Separator and Accelerator (ISAC) facility at TRIUMF. The ions are delivered to our experiment at the rate of 4$\times$10$^7$$/$s to 2$\times$10$^9$$/$s, and we collect them online on a  zirconium foil neutralizer of area 19 $\times$ 12 mm$^2$ and of thickness 0.03 mm. Typically, we collect the ions  on the foil for 20 s before rotating the foil by 90$^{\text{o}}$ and electrically heating it for 1 s to release neutral francium atoms from the heated foil (with maximum efficiency of 30$\%$). We collect the released atoms first in a  MOT inside a coated glass cell (collection chamber), then we transfer the atoms  to another MOT inside a stainless steel vacuum chamber (science chamber, with maximum transfer efficiency of 20$\%$ for this work), by pushing them with a pulse of laser light resonant with the D2 line in francium at 718 nm \cite{Zhang2016}. The MOT in the science chamber is located at 0.7 m, directly below the MOT in the collection chamber. We operate both  MOTs  on the D2 line of francium and they share two Ti:Sapphire lasers. We use one laser (MSquared SolsTIS) for trapping, and we use the other laser (Coherent 899-21)  for re-pumping the atoms. We maintain a pressure of $\approx$2 $\times$10$^{-10}$ Torr in the science chamber. A detailed description of the francium trapping facility (FTF) can be found in Refs. \cite{1748-0221-8-12-P12006,Zhang2016}. We can operate our apparatus with a range of isotopes ($^{206-213,221}$Fr) by adjusting our trap and re-pump laser frequencies, and requesting a specific isotope from ISAC \cite{PhysRevA.90.052502}.

We perform two-photon spectroscopy using  atoms confined in the MOT in the science chamber. We use a third Ti:Sapphire laser (MSquared SolsTIS) at 1012 nm as our spectroscopy laser, in order to excite the $7s(F=U)\rightarrow8s(F=U)$ transition, where $F=U(L)$ is the upper(lower) hyperfine manifold of the $s$ levels.

We lock  the frequencies of all  three lasers to a stabilized HeNe laser (Melles-Griot 05-STP-901) using  a computer-controlled feedback system \cite{:/content/aip/journal/rsi/69/11/10.1063/1.1149171}.

The linearly polarized spectroscopy laser beam of 350 mW  power is focused to  1/$e^2$ (intensity) diameter of 0.015 cm, using an achromatic lens of 30 cm focal length. The lens is mounted on a translational stage to fine-tune the overlap of the laser beam with the atom cloud. In order to increase the average intensity of the spectroscopy beam across the atom cloud, the beam is re-collimated and re-focused back on itself in a double-pass scheme using a second  30 cm focal length lens and a mirror. An optical isolator (LINOS, FI-980-TI) is necessary to reduce optical feedback into the laser.

We apply a frequency offset between the  beam pick off for locking and the spectroscopy laser beam directed at the atom cloud by using an acousto-optic modulator (AOM)  in double-pass configuration. 
We ramp the offset  by 18.86 MHz over 12 s (a 37.72 MHz scan across the 7s$\rightarrow$8s resonance).

\begin{figure}
    \center
    \includegraphics[width=.36\textwidth]{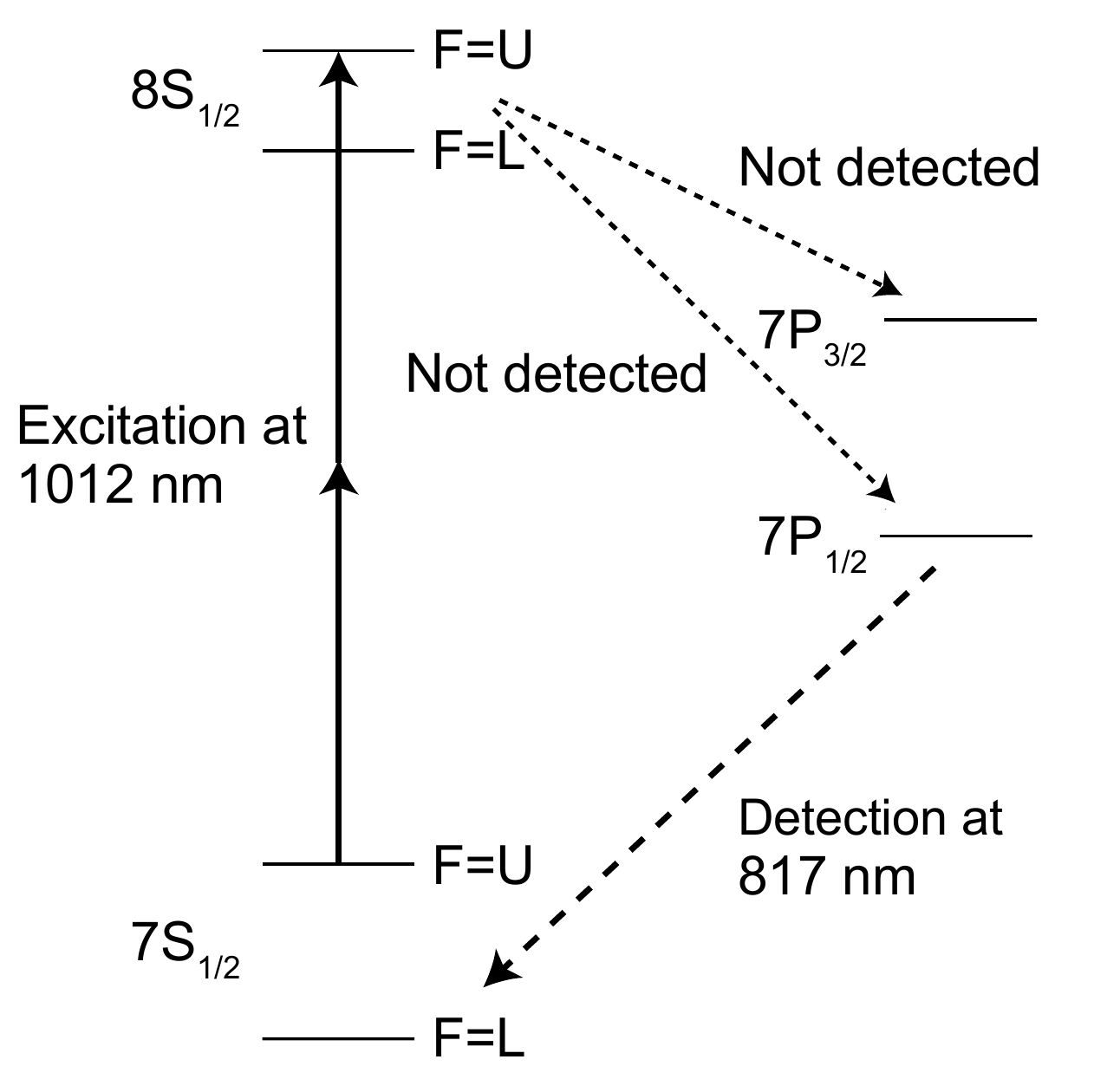}
    \caption{Energy level diagram for francium with relevant transitions. Atoms in the 7$s$ state are excited to the 8$s$ state with two 1012 nm spectroscopy laser photons (solid arrows). The spontaneous decay (dashed arrows) via the 7$p_{1/2}$ level is detected at 817 nm. This figure is not to scale. }
    \label{fig:level}
\end{figure}

We detect the resonance of the $7s\rightarrow8s$ transition by collecting 817 nm photons resulting from the decay of atoms from the $7p_{1/2}$ state to the $7s$ ground state (D1 line of francium and about 100 nm away from the the D2 line as shown in Fig.\ref{fig:level}).

We direct the 817 nm photons onto a  photomultiplier tube (PMT, Hamamatsu H7422 operated in  photon counting mode) using a  double-relay optical system. To reduce background counts from the trap beams at 718 nm, we place an edge filter (Semrock LP02-785RU) and a longpass coloured glass filter (Thorlabs FGL780M) in front of the PMT.  The 718 nm light scattered by the MOT does not contribute significantly to our background counts.
 We save PMT data as a function of time for later analysis. The beginning of the offset frequency scan and the beginning of PMT data collection is synchronized using a  digital trigger. During the scans, the trap light is cycled on and off with a 2 ms period (50\% duty cycle) and with an extinction ratio of 1000:1, while the repumper and spectroscopy light remain on continuously. We collect data when the trap light is off to suppress the ac Stark shift that it produces, as well as to minimize background counts. During each MOT collection-transfer cycle, we perform a single offset frequency scan of the spectroscopy laser. 

\section{\label{sec:level1}Results}\label{Results}

\begin{figure}
  \centering
   \includegraphics[width=0.46\textwidth,keepaspectratio]{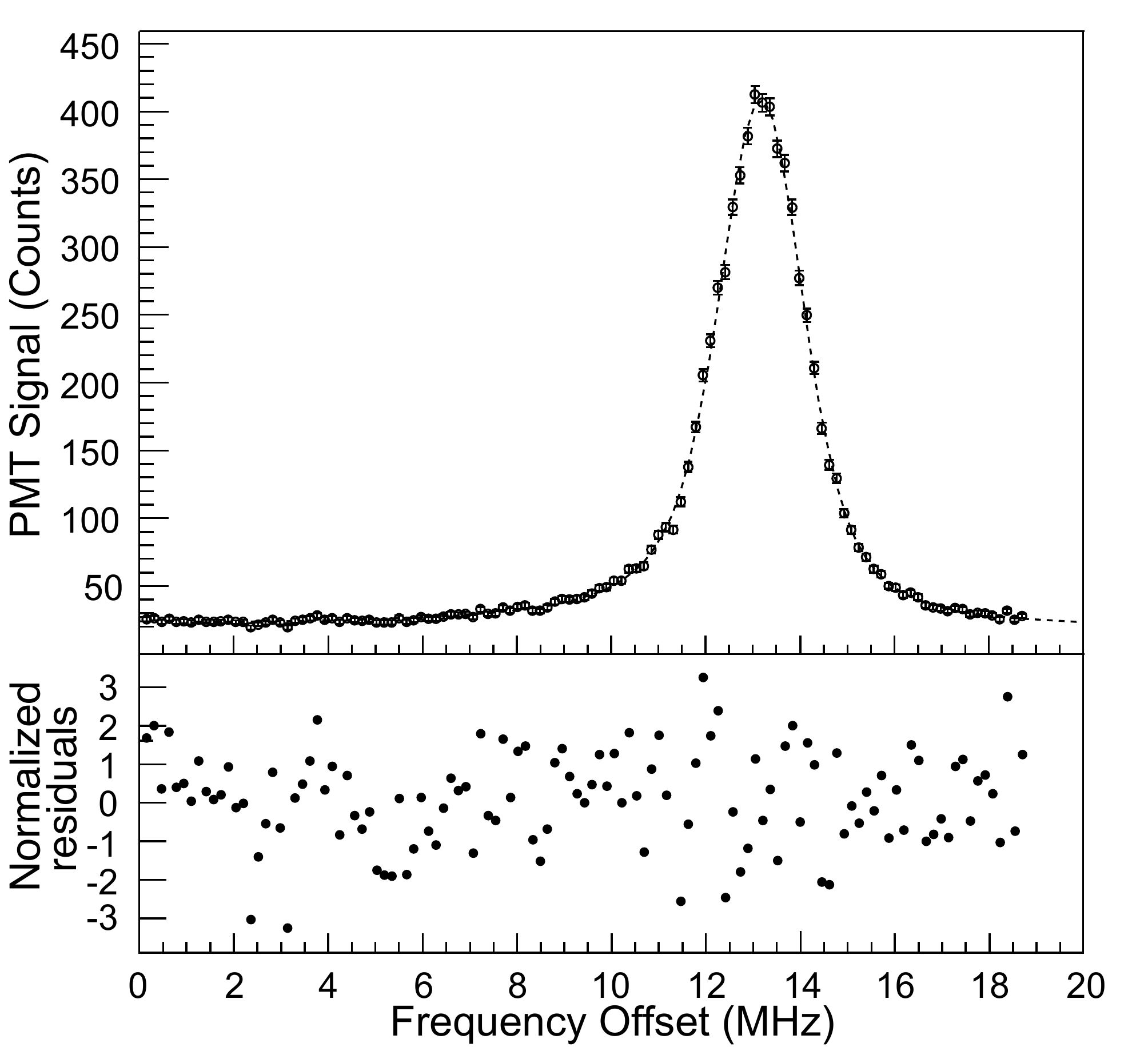}
  \caption{Two-photon spectroscopy data for the $7s(F=5) \rightarrow8s(F=5)$ transition in $^{211}$Fr. The vertical axis shows PMT counts  of 817 nm photons. The frequency scan starts at the zero of the horizontal axis.  The dashed line is a fit to the data (see text). The bottom plot shows the normalized residuals of the fit.}

\label{fig:awesome_image}
\end{figure}

The hyperfine interaction  splits the $s$ states into two hyperfine levels. We measure transition frequencies from the upper hyperfine level of the  $7s$ ground state to the upper hyperfine level of the  $8s$ excited state (Fig. \ref{fig:level}) in five different isotopes of francium:  $^{208}$Fr (radioactive half-life T$_{1/2}$ = 59 s), $^{209}$Fr (50 s), $^{210}$Fr (192 s), $^{211}$Fr (186 s) and $^{213}$Fr (35 s). Fig. \ref{fig:awesome_image} shows typical 817 nm fluorescence  for  scan of the two photon excitation in the isotope $^{211}$Fr. 10 scans of 12 s duration each are used to generate this plot. The separation between bins is 157 kHz.  


In order to locate the fluorescence peak, we fit the data to a product of an exponentially decaying function and a Voigt function using  a computer program that utilizes the ROOT analysis framework.  The exponential decay takes into account the exponential rate of loss of atoms from the trap (the 1/$e$ lifetime of the atoms in the trap can be as long as 14 $\pm$ 3 s) during the laser scans.
 The program uses the ``MINUIT" function minimization package to find the best parameters that minimize the $\chi^2$ of the fit \cite{BRUN199781}. We float the following fit parameters: peak position, background level, width (Lorentzian and Gaussian), $1/e$ decay constant, and peak height. Over the five isotopes, the $\chi^2$/(degree of freedom) for the fits varies from 1.2 to 1.8.

The 7$s\rightarrow 8s$ transition energy $\nu_0$ is given by the relation $\nu_0=2\times(\nu_f+\nu_w)$, where $\nu_f$ is the offset frequency of the peak from the fit and $\nu_w$ is the  probe laser frequency at the beginning of a scan, i.e. for zero offset frequency, as measured with our wavemeter (Angstrom WS-U-10). The factor of two is due to the two-photon nature of the transition that we excite. 
 We correct the wavemeter reading by comparing it to a diode laser referenced by saturated absorption to the $5s(F=2)\rightarrow5p_{3/2}(F=3)$ transition of $^{87}$Rb \cite{Ye:96}.

Using 7$s\rightarrow 8s$  transition energies from our measurements and  previously published measurements of the hyperfine splittings of the $7s$ state of the   isotopes \cite{COC198566}, and the hyperfine splitting of the $8s$ state in $^{210}$Fr \cite{PhysRevA.59.195} and considering that the ratio of the hyperfine splittings of the $7s$ and the $8s$ states are same across the isotopes (because the difference in hyperfine anomaly is negligible for $s$ states with different quantum number), we determine the C.O.G. of the $7s\rightarrow 8s$ transition in the five isotopes. We obtain the  isotope shifts of the $7s\rightarrow8s$ transition by subtracting the C.O.G. of the transition in the isotopes  from the C.O.G. of the same transition in  $^{213}$Fr.

We find the 1$\sigma$ error in the peak position obtained from the fitting program to be less than one bin. We assign a conservative estimate of 157 kHz (frequency separation between bins) to  the error in the peak position obtained from the fit.

From our data we estimate the uncertainty in the peak position ($\nu_f$) obtained from the scans to be 1 MHz. For this analysis, we start each laser scan  with a similar laser frequency as measured by our wavemeter with 1 MHz resolution. We assign 2 MHz uncertainty to the isotope shift measurements due to non reproducibility of the scans.

The ac Stark shift due to the trap light at 718 nm is reduced by dimming the trap light by a factor of 1000 during data collection yielding a negligible contribution to the uncertainty.

The ac Stark shift of the 1012 nm light that we use to drive the  $7s\rightarrow8s$  transition was theoretically studied in Ref. \cite{PhysRevA.93.043407}. For our typical 1012 nm laser power of 350 mW and beam diameter of 150 $\mu$m, the estimated shift is $<$ 50 kHz. The laser power is typically stable at the $<$ 5\% level, and the error on our measurements due to this effect is negligible.

The energy levels involved in the $7s\rightarrow8s$ transition have similar $g$ factors and hence similar Zeeman effects. There is no linear shift in the measured transition frequency due to the magnetic field gradient of 10 G/cm of our MOT (this is due to  the $\Delta m_F= 0$  selection rule). The cold atom cloud has a diameter of about 1 mm and resides close to the zero of the magnetic field. We do not include any error or systematic shift on the the isotope shift measurements due to  magnetic fields.

We add all these errors in quadrature to estimate the uncertainty on our measurements of the transition frequencies of  the $7s(F=U) \rightarrow 8s(F=U)$ transition in the five isotopes.  In order to determine the error in our calculation of the C.O.G.,  we use the reported errors in the measurements of the hyperfine constants of the $7s$ and $8s$ states from Ref. \cite{COC198566} and Ref. \cite{PhysRevA.59.195}.
The results from our measurements and the isotope shifts in the D1 line of francium from Ref.  \cite{PhysRevA.90.052502} are shown in Table \ref{fig:awesome_table}. The isotope shifts in the D1 line are calculated for this analysis from data reported in Ref. \cite{PhysRevA.90.052502}, using  $^{213}$Fr  as the reference isotope.
\vspace*{-\baselineskip}
\begin{table}[H]
\caption{\label{tab:example} Transition frequencies and isotope shifts of the $7s\rightarrow 8s$ transition ($\delta\nu_{IS,SS}$)(this work), and isotopes shifts in the D1 line ($\delta\nu_{IS,D_{1}}$) based on  measurements reported in Ref. \cite{PhysRevA.90.052502}. Isotope shifts are calculated using $^{213}$Fr as the reference isotope. }
\begin{ruledtabular}
\begin{tabular}{ccccccc}

    \multirow{5}{*}{} &
      \multicolumn{3 }{c}{This work} &
      \multicolumn{3}{c}{Ref.  \cite{PhysRevA.90.052502}} \\
      \hline
      \Tstrut\Bstrut \\
   Isotope &\multicolumn{1}{p{1cm}}{\centering Nuclear spin \\ ($I$)}  & \multicolumn{1}{p{1.3cm}}{\centering {C.O.G. $7s\rightarrow8s$    \\ (cm$^{-1}$)}} & \multicolumn{1}{p{1cm}}{\centering $\delta\nu_{IS,SS}$  \\ (MHz)}&\multicolumn{1}{p{1cm}}{\centering $\delta\nu_{IS,D_{1}}$ \\ (MHz)}  \\
    \hline\Tstrut\Bstrut\\
    208 & 7   & 19732.58581(18) & -5123(6) &-6341(5)  \\

    209 & 9/2  & 19732.53758(15) & -3678(6) &-4563(4)  \\

   210 & 6  & 19732.52411(15) & -3274(6) &-4058(4)   \\

    211 & 9/2  & 19732.48021(15) & -1958(6) &-2431(4)   \\

      213 & 9/2  & 19732.41489(15) & 0 & 0 \\

\end{tabular}
\end{ruledtabular}
 \label{fig:awesome_table}
\end{table}\vspace*{-\baselineskip}

 \section{\label{sec:level1}King Plot analysis}\label{King Plot analysis}
\begin{figure*}[ht]
    \center
    \includegraphics[width=.67\textwidth, keepaspectratio]{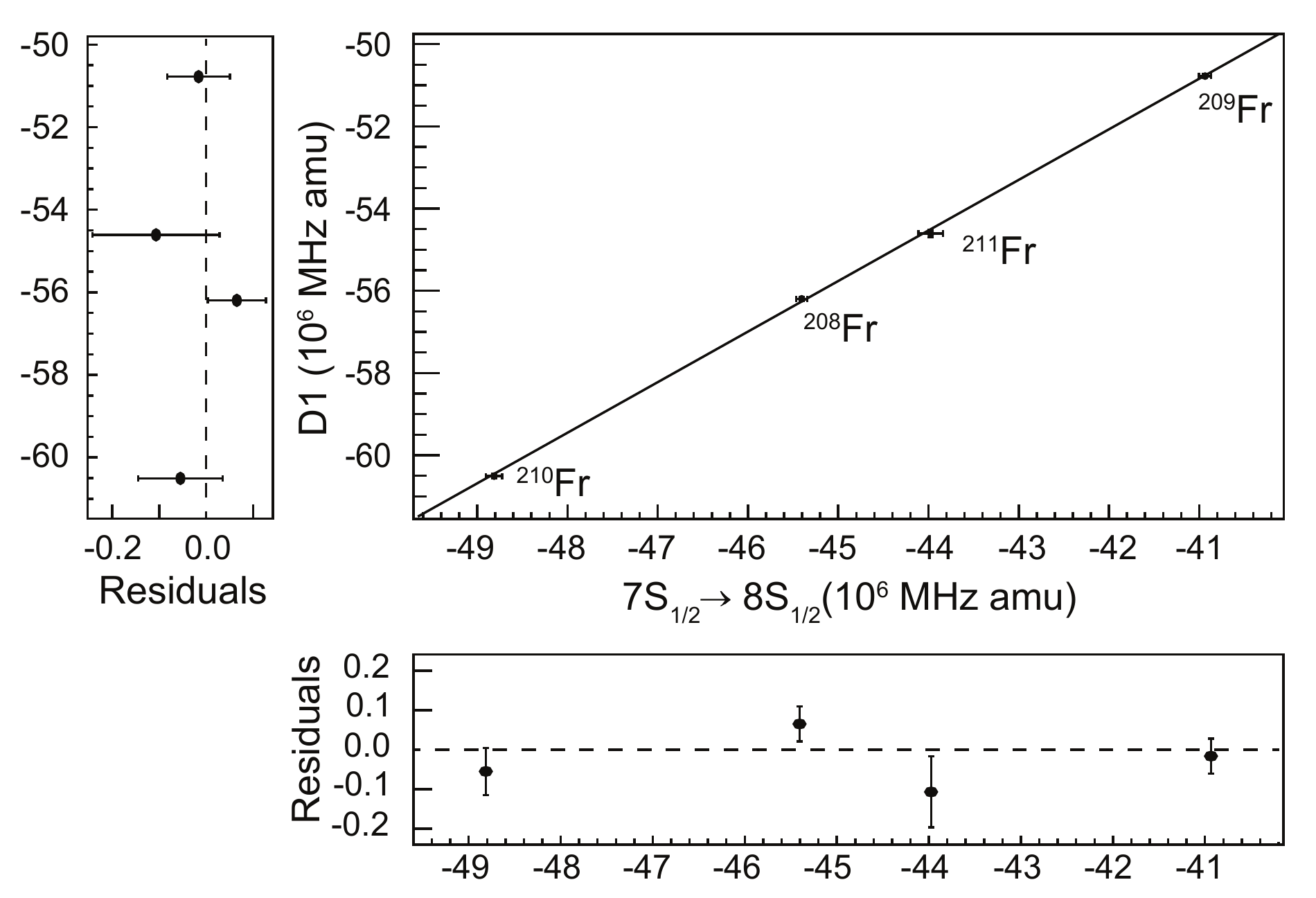}

    \caption{The modified isotope shift of the D1 line is plotted against the modified isotope shift of the $7s\rightarrow8s$ transition (following relationship \ref{eq:kps}) for the isotopes $^{208}$Fr, $^{209}$Fr, $^{210}$Fr and $^{211}$Fr relative to $^{213}$Fr.  The solid line is a fit to the data. The error bars are calculated from the errors reported in table \ref{fig:awesome_table} and the reported error in the masses of the isotopes \cite{nist}.}
    \label{fig:kp}
\end{figure*}

In order to perform the King Plot analysis, we plot the modified isotope shifts of the D1 line against the modified isotope shift of the $7s\rightarrow8s$ transition. This gives a straight line  \cite{King:63} according to the relationship
\begin{equation}
\begin{aligned}
\frac{M_AM_{A^{'}}}{M_A-M_{A^{'}}}\delta\nu_{IS,D1}={} &\frac{F_{D1}}{F_{SS}}\frac{M_AM_{A^{'}}}{M_A-M_{A^{'}}}\delta\nu_{IS,SS}+ \\ &(N_{D1}+\mathcal{S}_{D1})-\frac{F_{D1}}{F_{SS}}(N_{SS}+\mathcal{S}_{SS}),
  \end{aligned}
 \label{eq:kps}
\end{equation}
where $N_{D1}$($N_{SS}$), $\mathcal{S}_{D1}$($\mathcal{S}_{SS}$) and $F_{D1}$($F_{SS}$) are the normal mass shift, specific mass shift, and the field shift of the D1( $7s\rightarrow8s$) transition, with $M_A$ as the mass of the reference isotope.
The resulting King Plot is shown in Fig.\ref{fig:kp}.

 We fit the data obtained from the King Plot  to a straight line using a computer program that utilizes the ROOT analysis framework. The program minimizes  $\chi^2$ of the fit using ``MINUIT", taking into account  errors in both the horizontal and the vertical axes \cite{Press:2007:NRE:1403886}. Fig. \ref{fig:kp} shows the fitted straight line to the data. We find the value of $\chi^2$/(degree of freedom) from the fit to be 0.54. This corresponds to a P value of 0.58 for our straight line fit to the data. The slope  is equal to the ratio of the field shift constants of the D1 transition and of the $7s\rightarrow8s$ transition according to Eq. \ref{eq:kps}. This  represents the ratio of the change in electron densities at the nucleus during the corresponding transitions. Since an $8s$ electron has a larger probability density at the nucleus compared to a $7p_{1/2}$ electron, the ratio of the field shift constants is expected to be greater than 1. From the fit we find $\frac{F_{D1}}{F_{SS}}$  = 1.230 $\pm$ 0.019. We compare this result to the theoretical  value of ${\mathcal{R}}$ (Eq. \ref{eqR}) of 1.234 $\pm$ 0.010 from Table \ref{tab:1} and find excellent agreement. From  the intercept of the straight line, we find  $(N_{D1}+\mathcal{S}_{D1})-\frac{F_{D1}}{F_{SS}}(N_{SS}+\mathcal{S}_{SS})$ = ($-$0.41 $\pm$ 0.85)$\times$10$^{6}$ MHz amu. The errors reported here for the slope and the intercept   are the 1$\sigma$ errors obtained from the fit. The normal mass shift constant for the D1 transition is $N_{D1}$= 201 GHz amu, and the normal mass shift constant for the $7s\rightarrow8s$ transition is   $N_{SS}$= 325 GHz amu. From this we find : $\mathcal{S}_{D1}-\frac{F_{D1}}{F_{SS}}\mathcal{S}_{SS}$ = $-$214(847) GHz amu.
 \section{\label{sec:level2}Conclusion}\label{Conclusion}
  
We have observed the electric dipole forbidden $7s\rightarrow8s$ transition using two-photon excitation in five different isotopes of francium towards our efforts to perform APNC experiment in this rare alkali atom. Combining our measurements with previous measurements of the $7s\rightarrow7p_{1/2}$ transition we have performed a King Plot analysis and extracted the ratio of field shift constants. Our measurements provide benchmarks for theoretical calculations necessary to interpret results of future APNC experiments. Towards this we find excellent agreement between \textit{ab initio} theory and experiment for the ratio of field shift constants.        
 
  \section{\label{sec:level2}Acknowledgements}\label{Acknowledgements}
  
  We thank the ISAC staff at TRIUMF for developing the francium beam and Mikhail Kozlov for helpful discussions. The francium experiment is operated with NSERC (Canada) and NSF PHY-1307536 (USA) grants and was established with equipment funds by DOE (USA) and NSERC (Canada). TRIUMF receives federal funding through a contribution agreement with the National Research Council of Canada. M.S.S. is supported by NSF grant PHY-1620687 (USA) and the UNSW group by the Australian Research Council. M.S.S. thanks the School of Physics at UNSW, Sydney, Australia for hospitality and acknowledges support from the Gordon Godfrey Fellowship program, UNSW. S.A. acknowledges support from Fulbright Canada, and E.G. from CONACYT (Mexico). A.C.dH and M.J.K. were supported in part via the University of Manitoba GETS program.

\bibliography{Francium7s8s}

\end{document}